\documentclass[manuscript]{aastex}	
		\usepackage{epsfig}
		\begin{document}

		\title{$K$-minus Estimator Approach to Large Scale Structure}		
		\author{M. Martinis and M. \v So\v si\' c}		\affil{Rudjer Bo\v skovi\' c Institute, Zagreb, Croatia }		\email{martinis@irb.hr}

\begin{abstract} Self similar 3D distributions of point-particles, with a given quasifractal dimension $D$, were generated on a Menger sponge model and then compared with \textit{2dfGRS} and \textit{Virgo project} data \footnote{www.mso.anu.edu.au/2dFGRS/, www.mpa-garching.mpg.de/Virgo/}. Using the principle of local knowledge, it is argued that in a finite volume of space only the two-point minus estimator is acceptable in the correlation analysis of self similar spatial distributions. In this sense, we have simplified the Pietronero-Labini correlative analysis by defining a $K$-minus estimator, which when applied to 2dfGRS data revealed the quasifractal dimension $D\approx 2$\ , as expected. In our approach the $K$-minus estimator is used only locally. Dimensions between $D = 1$ and $D = 1.7$, as suggested by the standard $\xi (r)$ analysis, were found to be fallacy of the method. In order to visualize spatial quasifractal objects, we created a small software program called \textit{RoPo} (''Rotate Points''). This program illustrates and manifests local correlative analysis in which the visual inspection emerged as a first step and a key part of the method. Finally, we discuss importance and perspective of the visual inspection on available real and simulated distributions. It is also argued that results of contemporary cosmological simulations do not faithfully represent real data, as they show a formation of ever increasing collapsars. We consent that 2dfGRS data are reminiscent of some kind of underlying turbulence like effects in action.	
\end{abstract}


\section{Introduction}

 A statistical approach to the analysis of galaxy and cluster correlations (like in \cite{Gab05}) has led to the surprising result that they may be fractal up to the limits of the available catalogs. This  so called fractal conjecture is based on the analysis of two-point correlation function, presently the main statistical tool in the observational cosmology. From this point of view and together with the emersion of modern redshift surveys, the galaxy structures appear highly irregular and self-similar with fractal dimension $D \sim 2$ up to the deepest scales probed so far (1000$h^{-1}$Mpc) and maybe even more as indicated from the interpretation of the number counts as in \cite{Lab98}. 

Clearly and obviously, space is highly structured. However, a meaningful explanation for this large scale structure (LSS) is still lacking, and this remains the key question in contemporary cosmology. Its resolution seems crucial for possible extension of general relativity to the quantum domain and so on. 

In the following, respecting the fractal conjecture, we comprehend and repeat the correlation analysis of available data. We use the $K$-minus estimator as a tool in spatial statistics when  dealing with selfsimilar structures. 

As usual, the three-dimensional distribution of galaxies and   generated quasifractal objects are represented by a set of points in empty space. Their density is given by the  sum of delta functions $n(\vec r) = \sum\limits_{i}{\delta (\vec r - \vec r_i )}$ \ so reducing all properties of these objects to simple material (mathematical) points, defined only by their positions in space. Of course, it is always possible to add  multiple attributes (mass, ``colour", orientation, ...) to these points \footnote{In a perspective even thousands of simple or extended attributes can sit in the background, organized in some database model. Variety of metrics can be conceived, etc. Connections also may have attributes, we will call them ``predicates" or ``relations", but this is another subject not to be discussed now.}. 

Our starting relation is $N(r) \sim r^D$, known as a ``mass-length" relation (\citet{Bav98}). We imagine a sphere of radius $r$ around every point of a quasifractal object and count all the points inside the sphere to get $$ N(r) = \sum_{i}\theta (r - r_{i}) = N_{trend}(r) + N_{fluc}(r)$$ The ``mass-length'' relation refers to the behaviour of $N_{trend}$. In further discussion $\sim$ will be used in a sense of trend, or mean. 

A quasifractal object is made up of a large but finite number of points. It is usually only approximately self-similar at some scale range. The scale is necessarily restricted from above by the size of the sample, and from below by the size of the galaxy itself (for more details see \citet{Bav98}). Such a scale bounded object  will be referred to as  quasifractal to distinct it from an exact fractal which is composed of an infinite number of points and is exactly self similar at all scales, in other words, it is infinitely large in extend and infinitely structured \footnote{reader may try \textit{GNU XaoS} software}. The ``mass-length" relation of a   quasifractal object has  a power law trend only in some range of scales.  

An interesting question arises now, how these random three-dimensional quasifractal objects really look like. To answer this question, we created a small program for generating various such objects using the Menger sponge model (MSM)  \footnote{program is named ``menger.for" and is available upon request from {\it martinis@irb.hr} ; it does 5 or 6 selfsimilar recursive iterations depending either on the choice of $N$ and $1/r$, or on the available number of sample points}. The fractal dimension is determined by the formula $Nr^D = 1$, as in \citet{Bav98} so that we need only to give two numbers, $N$ and $1/r$, respectively. 

By using an uniform random number generator  for position coordinates, we also generated an ``uniform gas" of particles. In our discussion, it is regarded as a quasifractal object too, with  fractal dimension $D=3$. Although an uniform gas is a trivial quasifractal object, it is important in the correlation analysis for it has no imposed or presumed inner structure. As such, it is suitable  for basic testing  of various two point estimators. 

Our correlation analysis is based on the simplification of the Pietronero--Labini $\Gamma (r) $ and $\Gamma ^* (r)$ functions as presented in \cite{Lab97}:

$$\Gamma (r) = {1 \over N}\sum\limits_{i = 1}^N {{1 \over {4\pi r^2 \Delta r}}\int\limits_r^{r + \Delta r} {n(\vec r_i + \vec r')d\vec r'} } \sim r^{D - 3} $$
$$\Gamma ^* (r) = {3 \over {4\pi r^3 }}\int\limits_0^r {4\pi r'^2 } \Gamma (r')dr' \sim r^{D - 3} $$

We make use only of the integral kernel of the $\Gamma (r)$ function parameterized by $\Delta r$ and define our $K$ function as:

$$ K(r, \Delta r) =  \sum\limits_{i = 1}^N K_{\vec r_i}(r, \Delta r)  =   \sum\limits_{i = 1}^N \int\limits_r^{r + \Delta r} {n(\vec r_i + \vec r')} d\vec r' \sim r^{D - 1} $$

and $r < \left| {\vec r'} \right| < r + \Delta r$. This is our main formula. It can be easily generalized to the multidimensional case. $K(r, \Delta r) $ and $K_{\vec r_i}(r, \Delta r)$ are $\sim r^{D - 1}$ (trend) in the case of quasifractal. 

Simply as it is, $ K_{i}(r, \Delta r) $ counts the number of points falling into the spherical shell $(r, r + \Delta r)$ around some starting point $\vec{r}_i$ (the origin). We sum over all such starting points that can be found in the volume of the sample. In fact $ K(r, \Delta r) $ counts the number of pairs, or the number of connections between the points. Any correlation analysis basically count and classify connections between points. Obviously, $K$ is the simplest way of doing that. 

Pair connections and their possible attributes  are usually classified by length only, though we may conceive many other classification criteria.

No additional statistical or probabilistic interpretation of $K(r, \Delta r)$ is needed, in particular there is no need for a normalization, sample defined multipliers, or for the problematic factors $1/{r^2}$ and $1/{\Delta r}$ as in $\Gamma (r) $ and $\Gamma ^* (r) $ functions, respectively.  We are interested only in the trend of $K$, expected to be $\sim r^{D - 1}$ and in the determination of $D$ -- we quit any additional statistical interpretation of correlation analysis apart from classifying connections. This simplification is found to be very convenient in the correlation analysis of highly irregular selfsimilar spatial distributions. 

In most of the present literature it is debated how to define average  densities of irregular distributions. Here, we  make a philosophical detachment and debate about connections (pair links) only, avoiding density as much as possible. It is a well known fact, that an ``average density" cannot be defined for a selfsimilar (fractal) distribution as  it depends on the scale and the sample volume. In the case of a mathematical fractal, the average density is zero (``dominated by space") and the only exception is an uniform gas ($D=3$). Therefore and furthermore, fluctuations are also a problem -- fluctuations around what? There is neither a reference object nor a reference scale - all scales are relevant. We generally avoid using terms like homogeneity and isotropy (\cite{Lab97}), and anything derived from them. 

Estimator is a particular implementation of a correlation function - let's say numerical procedure. Examining formula (1), we recognize immediately the necessity that all spherical shells remain entirely in the volume of the sample. This means that some points in the sample near the edge cannot be used as starting points for a given shell ($r, \Delta r$) - there are shells not fully in the sample. From this observation, we construct the so called minus estimator $\hat K(r,\Delta r)$.

For a reasonably small $\Delta r$, the trend of the $K(r, \Delta r) $ function behaves like a derivative of the original mass-length relation $N(r) \sim r^D$. Of course, we didn't specify this case in the theory, but in practice this is the case when  constructing histograms.
 
We notice that all information $\Gamma (r) $, $\Gamma ^*(r) $ and $\xi (r) $ can offer are already fully contained in $K(r, \Delta r)$. For example, $\xi (r) = {{\Gamma (r)} \over {n_{average} }} - 1$, with $\Gamma (r)$ as a key part, and $-1$ as an unimportant part (``noise"), has a trend $ \sim r^{D - 3}$ only in the region where $-1$ is small compared to the first part (extended discussion in \cite{Bar05, Bar98, Bar99}). Among many other flaws, $\xi (r)$ has also a problem with the definition of an average density $n_{average}$ which cannot be properly defined in the case of quasifractal distributions. Nevertheless, $\xi (r)$ can be applied to continuous distributions with small (or very small) fluctuations of average density, such as dense fluids (or perhaps to cigarette smoke like distributions), but not to self-similar distributions. We wonder why majority of authors, in spite of problems, prefer to use $\xi (r)$.

\section{$K(r, \Delta r)$ as a minus estimator and construction of histograms}

A spatial pattern of galaxies may be virtually infinite in extent, but we can observe only part of it through a restricted window (``sample") which give rise to a sampling bias and censoring (missing data). When a spatial point distribution  is observed through a bounded window, edge effects hamper both the information about the properties of the pair  distance distribution and on the  function $K(r,\Delta r)$. In fact, the distance from a given starting  point to the next point of the sample  is censored by its distance to the boundary of the sample. Edge effects become obviously more severe  as the distance $r$ increases.

In a sample with $N$ points  there are $N(N-1)/2$ pair distances. From this set of pairs, we can use only a subset defined by  pairs of points with at least one point which is more than $d$ (``maximum range") away from the nearest sample boundary.

In spatial statistics, one traditionally uses the edge-corrected estimators which are weighted empirical distributions of the observed pair distances. The simplest approach is the ``border method", where attention is restricted  to those starting points lying more than $d$ units away from the boundary.  All pair distances up to $d$ are observed without censoring. This approach is often justified by appealing to the ``local knowledge principle" of mathematical morphology stating that we can know only what is in the sample, and draw conclusions only from that knowledge. However, the border method usually discards much of  data points, especially in three dimensions that can be unacceptably wasteful when estimating $K$.

If we take into account the local knowledge principle, thoroughly discussed by Baddeley et al. \cite{Bad97}, we can  avoid making any homogeneity or isotropy presumption.  The use of $K(r, \Delta r)$ as a minus estimator (``the border method estimator") seems unavoidable on quasifractal samples. Also, this estimator is very simple to implement. An observer must first make a shift $d$  from the edge  of the sample in order to choose the first point of  the pair. The second point can be anywhere in the sample, but no more than $d$ units away form the first point. 

The sub-sample formed by  first points is, of course, smaller than the whole sample, and there is a compromise between the number of initially chosen points and the range $d$ that can be varied.

Edge effects are easily recognized on histograms, by decline of the graph, as soon as we approach the boundary of the sample (see Fig. 4.). There may be numerous histogram construction schemes, for example step of the histogram may be $\Delta r$ or a fraction of it for a smoother trend and so on. In calculations, we truncated decimals to get the index immediately, as in an appended procedure (see c source code in appendix) which is almost trivial, especially compared to other estimators. 

Generally, we avoid any edge corrections in treating apparently (or conjectured) selfsimilar or fundamentally irregular distributions. We simply don't know and cannot know what is outside of the sample volume where there may exist strong clustering of galaxies as well as big voids. In fact, only information that are contained in the sample volume should be used. Any estimator using weighting schemes or other kind of edge corrections necessarily puts in the calculation additional information that do not exist, and have no justification particularly in the case of random fractals or extreme clustering, making  the calculating results  more or less arbitrary. 

It will be later argumented that even if we use the minus estimators, as presently the best tools known to us, we cannot completely rely on the results of correlation analysis, which means that we cannot and should not draw any firm conclusions from these results, perhaps only to answer the question whether or not the observed distribution  shows self-similarity in some scale range. This answer is given by reading histograms only, with no further conclusions.

An example of a good application could be the mutual comparison of (slightly) different distributions. The correlation analysis yields a correlation histogram, which is a highly reduced and abstracted, imperfect and incomplete information about the distribution. Full information is, of course, given by specifying all points positions and their properties. In that sense, two different distributions may have the same histogram, but they are not even close in appearance. It is clear now that visual inspection must be also a key and unavoidable part of the correlation analysis. For this reason, we should first visually examine a given distribution, before any calculation, and see if it is appropriate for applying $\hat K(r, \Delta r)$ which means that the chosen  sample has no hollow in the middle where we search for  first points, etc.. A histogram is fair (trend can be recognized) even for a small number of points. We call this ``$N(N-1)$" effect. 

A \textit{special and rare}, unlike feature of a spatial point distribution is its self-similarity with a nontrivial quasifractal dimension $D<3$\ . If this feature exists, then there must be a mechanism or reason responsible for generating and supporting this fractal clustering (fractalisation with stable self-similarity) and not some other conceivable kind of clustering. Cases are just few: gas, collapse, fractal, then their combinations, and then something neither of these.

For a self-similar distribution, we expect a histogram trend to follow the power-law behaviour. The standard straight line adjustment using the least square method in a log-log scaled diagram, gives according to the $K(r,\Delta r) \sim r^{D - 1}$ formula, a power-law exponent, interpreted as $D-1$. The adjustment to other functional forms may be also conceived, mainly to parabola, to distinguish collapsars from a foam.

The traditional correlation analysis  uses all sample points equally and simultaneously, and in this sense, it is ``global". Our approach is basically ``local", it implies the use of sub-samples, even edges needn't to be exactly defined, although we should take care of the approximate range $d$. Any general conclusion about the whole sample should be avoided. Instead, we examine a correlative histogram around each point $\vec r_i $ determined by $\hat K_{\vec r_i } (r,\Delta r)$ and casually parameterized by $\Delta r$. We know from experience that this histogram is changing quickly from one point to another - so that giving too much weight to the correlative analysis is not necessary nor recommended. As an illustration our local correlative analysis is implemented in the \textit{RoPo} program together with the possibility of local sampling.

In most cases we can choose one maximal (cubical) sub-sample, and then discuss correlative analysis of the sample, not the analysis around each point. The calculation is demanding because of the great number of connecting pairs, so in order to limit calculations to a reasonable time span of few seconds, we used the Monte Carlo sampling procedure. It has not been proved yet, but it appears that a preliminary trend can be obtained already within the first (mili)seconds.

The advantages of $\hat K(r,\Delta r)$ minus estimator are the following: the histogram is always positive, it has a rising trend like parabolas ($\sim r^\nu$)\ ,  much easier perceived and interpreted than hyperbolas, which require log-log scaled diagrams.  Further advantages are: fast and simple  calculations compared to other estimators, and small hardware requirements. There is no need for additional distributions, normalizations, weighting schemes, etc. We believe it is the simplest method for testing self-similarity in spatial distributions of points.
                                                                                                                                                                            
A prominent and often mention case, relevant in cosmology, following for example \cite{Lab97, Lab98}, and others, is a quasifractal dimension $D=2$\ . On such distributions, our estimator gives a straight line, not only in log-log scaled diagram but also in the histogram itself. The interpretation of a histogram is very simple; we ask if the expected power-law behaviour $\sim r^\nu$ is a straight line $\nu =1$ (meaning $D=2$), or a square parabola $\nu =2$ (meaning $D=3$), or something in between \footnote{$D<1.8$ means a rather extreme clustering, we don't expect to find nothing such.}. Other cases are considered as deviations. Almost always a simple fast and rough visual estimate is sufficient to determine approximate quasifractal dimension $D$. It will be explained later, why more than that shouldn't be even asked.

A little bit of philosophy; to us scaling exists only after the correlation analysis and histogram inspection have been completed on a chosen sample. This depends obviously on the sample size and its properties. Then, if we  take another similar sample from the same distribution, and find no scaling, the question is whether the scaling exists or not? The only reasonable answer would be that locally on a particular sample, using a particular estimator, we see the scaling property. 

 On a gas-like distribution $\hat K(r,\Delta r)$ is almost exact (square parabola), and this is the only case. Therefore, we can use $\hat K(r,\Delta r)$ to determine whether or not and how much a certain spatial distribution is gas-like (homogenous), and this is probably the best use of $\hat K$.

\section{Compact samples and local analysis}

From a given large sample we pick up several sub-samples according to the following criteria. In order to avoid homogeneity and isotropy presumptions (anything with orientation should be carefully avoided), we require that a chosen sub-sample looks compact in its visual appearance. A criteria of minimal surface area to volume ratio can be established. An elongated sub-sample, a filament for example, has obviously a well defined orientation in space and it can fall along and/or across the sample creating the problem for the estimator. It  is rejected in our analysis because it calls for an isotropy presumption (equivalence of orientations). A spherical sample is the best example of a compact sample we are talking about. From the point of the calculation cost and speed, we prefer use of cubically shaped samples.  They do have a defined orientation, but in practice this is found to be unimportant.

 Elongated distributions such as 2dfGRS, the best available data, and SDSS, were analysed by first choosing  a number of cubical samples, which may or may not intersect each other, as it makes no difference in the final results. Then we found a trend for each of the cubic samples and took an average value of them, eventually rejecting some of the results that were too discursive for some reason \footnote{software \textit{Korela} has been made for processing a large number of samples in a sequence}. Initial samples were not of the same size (a greater redshift $z\sim r$, bigger sample), but we rescaled them to the unit cube. As we presumed self-similarity of 2dfGRS data, this rescaling made no difference, the structure should be the same on all scales. In this way, the points (data) deficiency at larger $z$ has also been avoided. We didn't try to correct any flaws of available data, nor did we select any particular distinguishing sub-set from them, by any criteria. It appears that  correcting flaws of data will not bring any significant change in the final results.

 \section{Special cases and application}

Let us now  make few comments on our analysis of generated self-similar distributions. In the case of an uniform gas, the error of the calculated quasifractal dimension was in the second decimal digit, or higher, while in the case of a quasifractal object with $D=2$, the error was in the first decimal digit. In order to obtain the result with an error in the second decimal digit, we had to generate and analyse about ten and more equivalent distributions generated with different random number seeds, and then take the average to discern imposed quasifractal dimension. Smaller quasifractal dimensions always mean stronger (more extreme) clustering for a given number of points in the sample and effects of the finite sample size and edge effects become more and more important. A lack of available starting points is also a problem. As we move away from $D=3$, any kind of estimator becomes less and less reliable. For $D=3$ all estimators may be used but for $D<2.6$ (estimated) we suggest only minus estimators. If there is any sense to talk about the error, it will be $D = 2.0 \pm 0.3$ in comparison with gas where we have $D = 2.988 \pm 0.064$, as tried on examples. 

From the above discussion, it has become clear that in a correlative histogram we cannot rely on one sample only and some kind of averaging over several equivalent samples is needed. For example, we can divide a big sample to several smaller subsamples. There are many possibilities. A series of evenly or randomly oriented samples can be chosen. Such calculation has been conducted on 2dfGRS data, to extract 259 samples on the first branch (some rejected)  finding  average $D=2.016$, and 195 samples on the second branch  finding average $D=1.976$. These findings are consistent with $D\approx 2.0$\ found  by other authors (for example Pietronero and Labini). 
 
We are not inclined to treat this result as a final truth, but only as an illustration.  
Furthermore, the range of correlation analysis is in this case quite small, we need  much bigger compact samples. From the point of view of our method, we can only draw  conclusions up to this small range.

The large-scale structure (LSS) of the universe has been heavily debated in the recent history without a firm conclusion. We believe that LSS, as well as interstellar gas, can be represented by random fractals. From 2dfGRS data, LSS is obviously not homogeneous \footnote{Upon request \textit{RoPo} freeware will be given; win32 for now, gcc port also exists. Software has been made for (serial) visualization of points distributions (simulation snapshots, or just a general point distributions, attributes also may be added -- colour for example). 2df.b data will be provided in a \textit{RoPo} .b format. Samples as big as 10 million points have been successfully visualized on a standard desktop computer. Main design objective was, of course, speed.}. The question still remains whether LSS is self-similar or not. Some theoretical aspects of possible self-similar LSS are discussed in \cite{Bar05}

\section*{Conclusion}

As big compact samples cannot be extracted from 2dfGRS, only the short range $K$-minus estimator correlation analysis is possible. A visual inspection remains the only alternative on a greater scale, and we think there is self-similarity on the entire available scale range and there is no visible crossing to homogeneity within 2dfGRS ($z=0.26$). 
We didn't treat less important effects like the Fingers-of-God, deficiency of points on greater $z$, errors in $z$ measurements, and so on, see \cite{Jon05}. A correlation estimator is a rough tool, as the  reader may have noticed from our previous discussion, so we don't consider this to be an important issue. It is the same with the visual inspection method - flaws of data may be ignored for the first try. 

Visual comparison with generated quasifractal objects and simulations data has been also  conducted. From it we decided by consensus that 2dfGRS data resembles fractal dimension between $D=2.1$ and $2.2$. What we see  is similar to a foam (or maybe ``drunk spider'' net), and  not  to various \textit{Virgo project} simulation distributions which are dominated by collapsars of different size. If those simulations do not provide something like 2dfGRS data, it means there is no explanation for this structure in the contemporary computational cosmology. There is nothing new in the theory either, although many alternatives and combinations of them have been suggested, MOND for example. From our perspective, some kind of turbulence should be incorporated in the simulation procedure (see picture of Crab Nebula's peripheral regions as an example). Of course, GTR should be a natural starting point.

Visual inspection is proposed here as an important part of the correlation analysis, but also as a sole method. This involves psychology and the knowledge on perception of density contrast seen on the screen, in our case bright and dark areas or groups of pixels on the black background. Perception is of course nonlinear. In addition, pixels may overlay, etc., so we don't have definite and precise answer to those issues \footnote{interesting subject, for example in automated medical imaging analysis, or maybe computer game industry}. But for the mere comparison of 2dfGRS and simulated data RoPo is sufficient, and we have firm conclusion that these distributions are  different.

The entire discussion about possible scale of transition to homogeneity should be postponed until at least one big \textit{compact} sample became available, and for that purpose 2dfGRS equipment could be reused. We think   that present data  do not show any transition to homogeneity. The fractal universe seems to be an alternative.

Why did we create the RoPo program? A simple reason was that  the  human brain and eyes perceive objects in motion much easier and deeper than just plain static pictures. We tried to emulate advantage of stereoscopic seeing (standing eyes but moving objects -- instead of standing objects but moving eyes as in nature), and it made a definite breakthrough in the discussion of 2dfGRS data morphology against the Virgo project (gravodinamical) simulation data morphology. In these simulations we clearly see two distinct classes of structures: collapsars and filaments. What is even more interesting, the filaments are almost all straight line alike, what we recognize as intersections of planes caused by gravity amplified inflicted plain wave perturbations. Those distributions are utterly dominated by collapsars, contrary to 2dfGRS where net-like structures dominate. 

To  resolve these problems more decidedly, we obviously need higher order correlation analysis. However at the moment there is no convincing way of defining three-points and other multi-points correlation functions in $K(r, \Delta r)$ paradigm. The question is whether this attempt would be worth doing it, because the two-point correlative analysis is already too approximate and can not produce a firm conclusion. It gives only an approximate and unreliable abstraction of properties of an underlying point process.

\appendix

\begin{figure}         
\includegraphics[scale=0.6, viewport=400 300  880 724,clip=true]{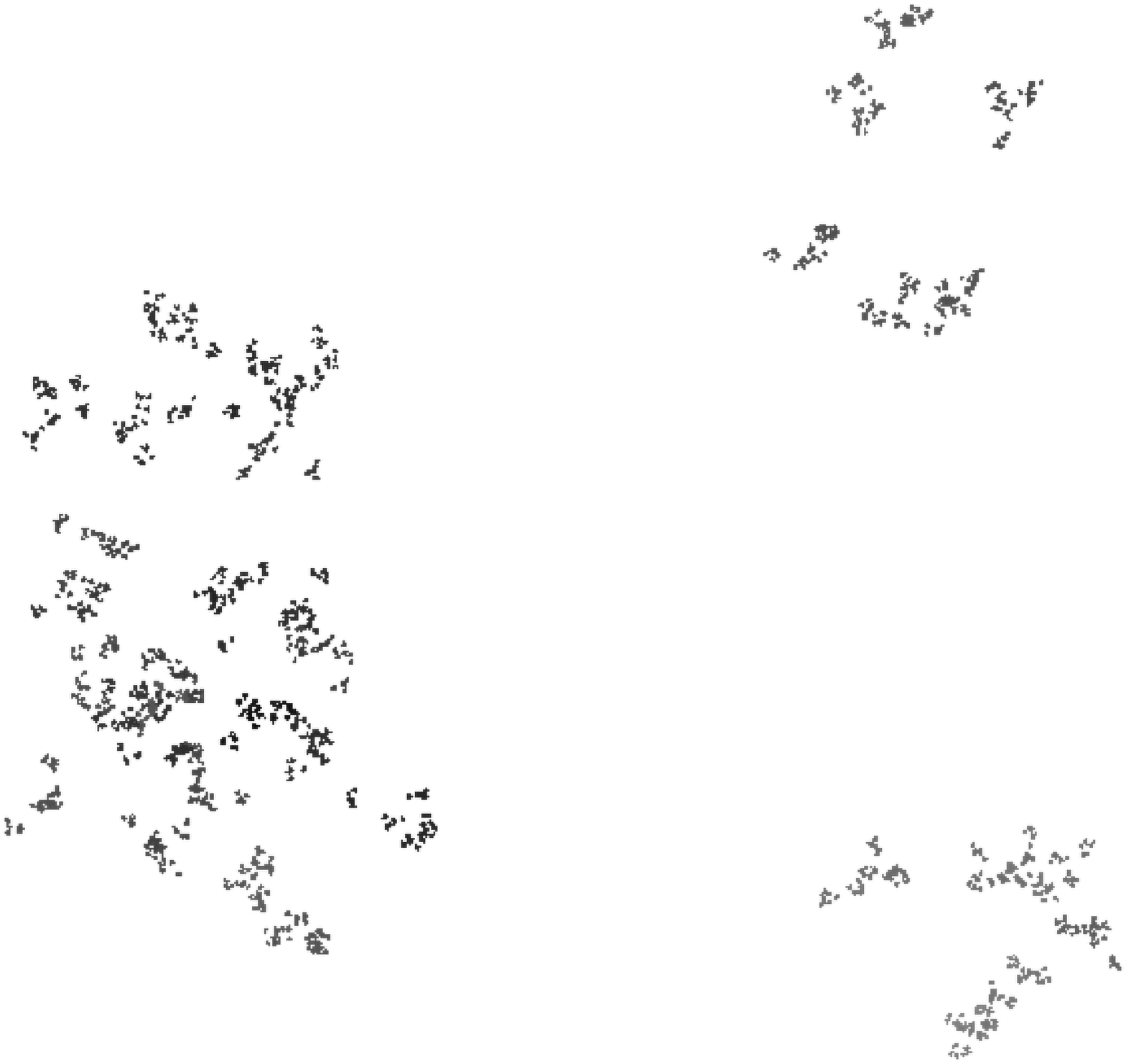}	
\caption{3D Quasifractal object with fractal dimension $D = 1.28$ containing 46656 points, as generated by a Menger sponge model. Distribution of points is dominated by large voids -- clustering is extremely strong. A lot of space had to be searched to find new points; see ``mass-length" relation. In the older literature one of the usual errors was comparing this with 2dfGRS data. Obviously they are not equal.}    
\end{figure}   							

\begin{figure}		\centering		\begin{tabular}{cc}
\epsfig{file=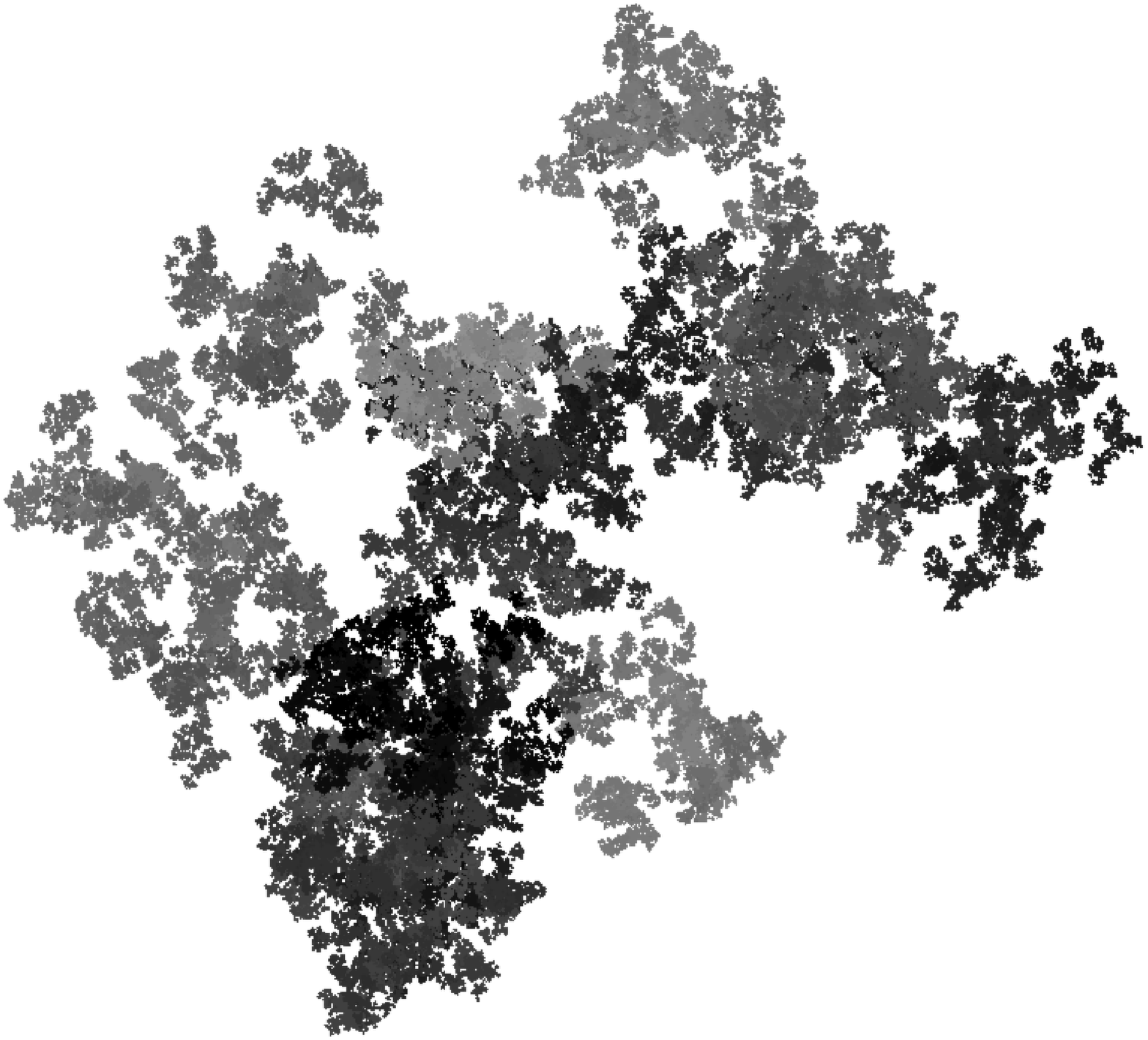,width=0.46\linewidth,viewport=400 300  880 724, clip=} &  \epsfig{file=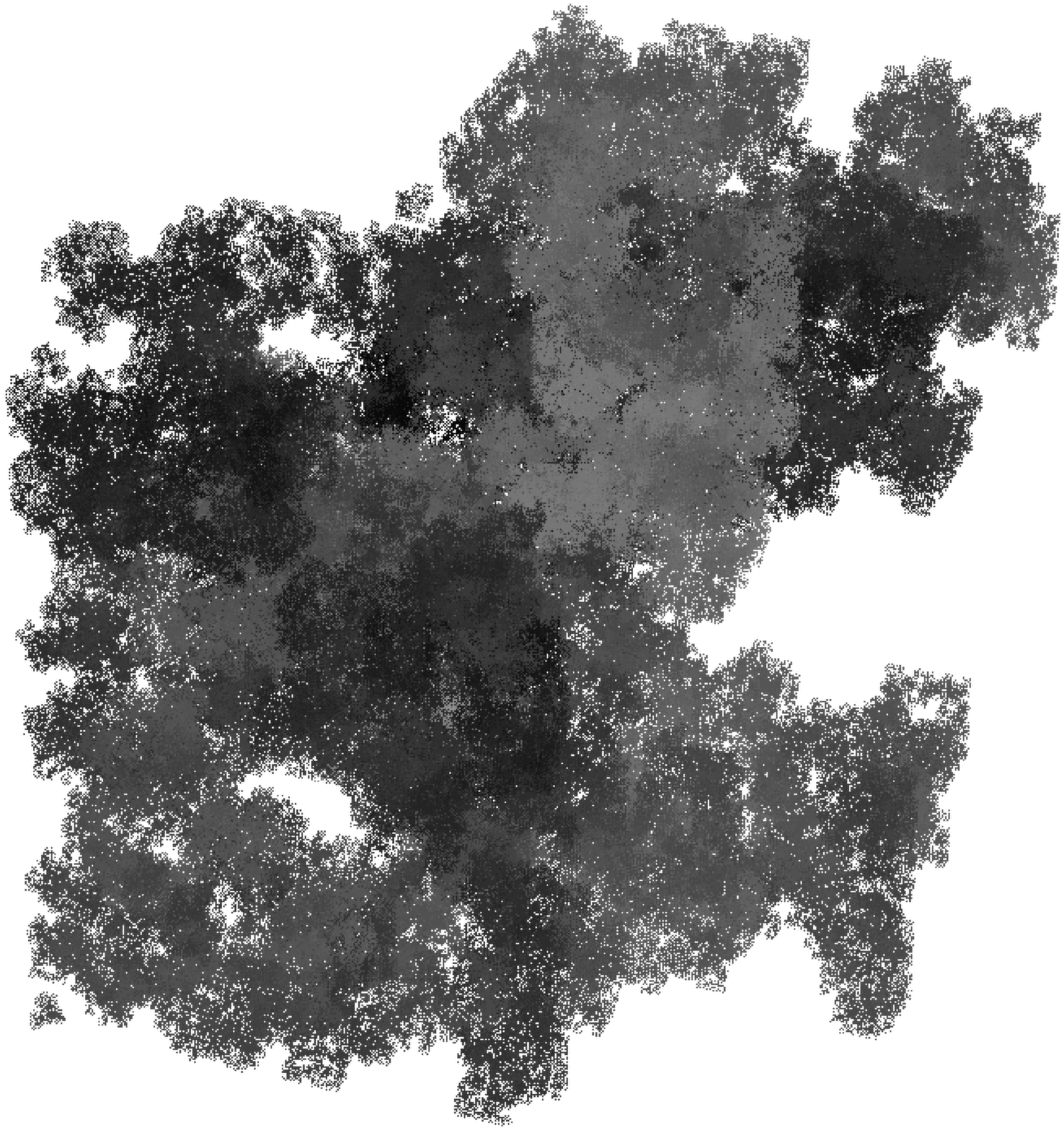,width=0.46\linewidth, viewport=400 300  880 724, clip=} \\
\epsfig{file=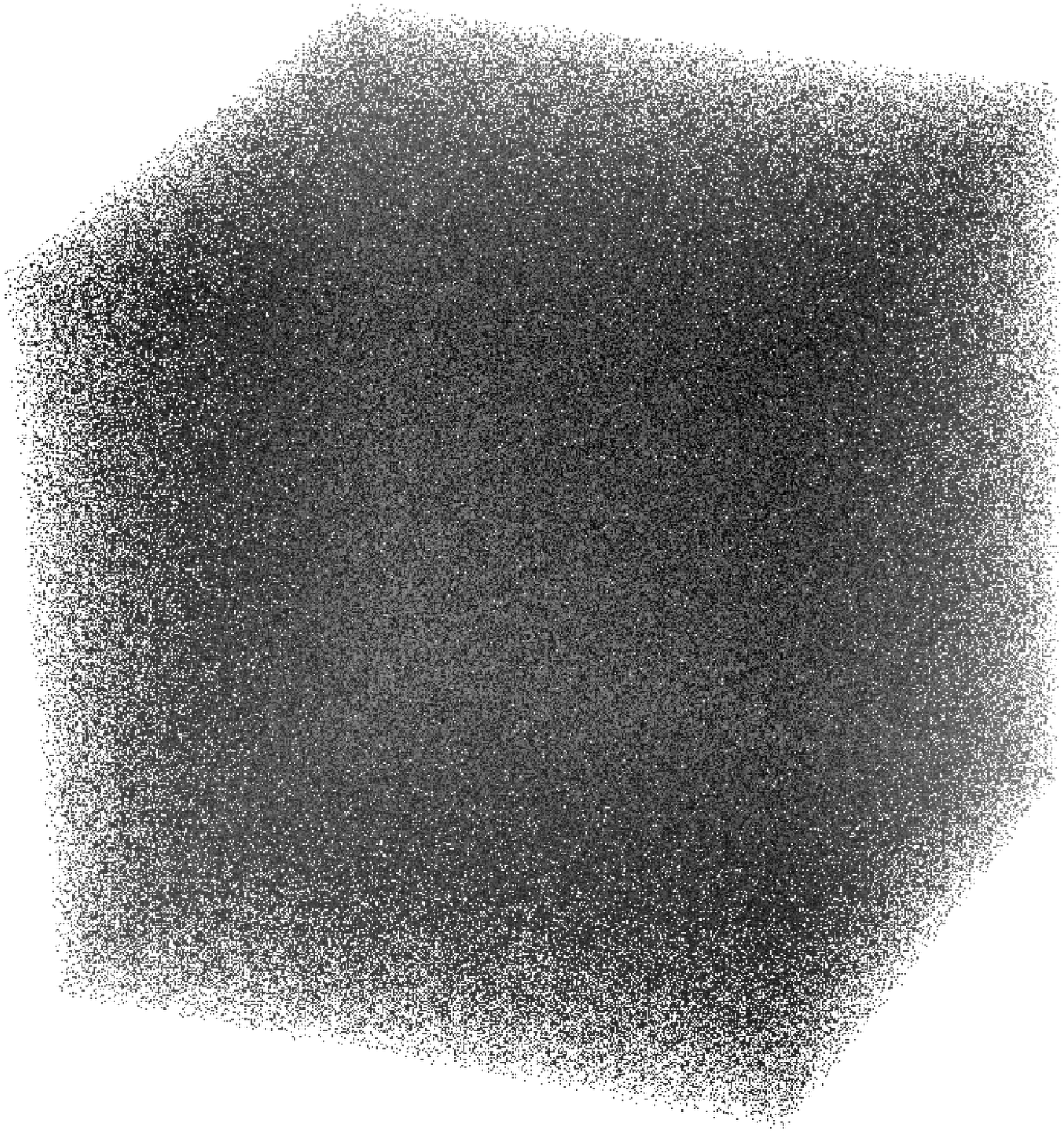,width=0.46\linewidth, viewport=400 300  880 724, clip=}		\end{tabular}
\caption{Quasifractal objects with fractal dimensions (a) $D=2.00$, (b) $D =  2.52$ and (c)  $D = 3.0$, respectively. Each picture has 1048576 points. We emphasize this number so that a reader can get impression of how self-similar clustering look like. All distributions are three-dimensional (note the shading) and they fit into the unit cube (menger generate by recursive and (quasi)random ``slicing'' from the unit cube). Distributions are rotated for a better view.}	
\end{figure}

\begin{figure}   \includegraphics[scale=1.2]{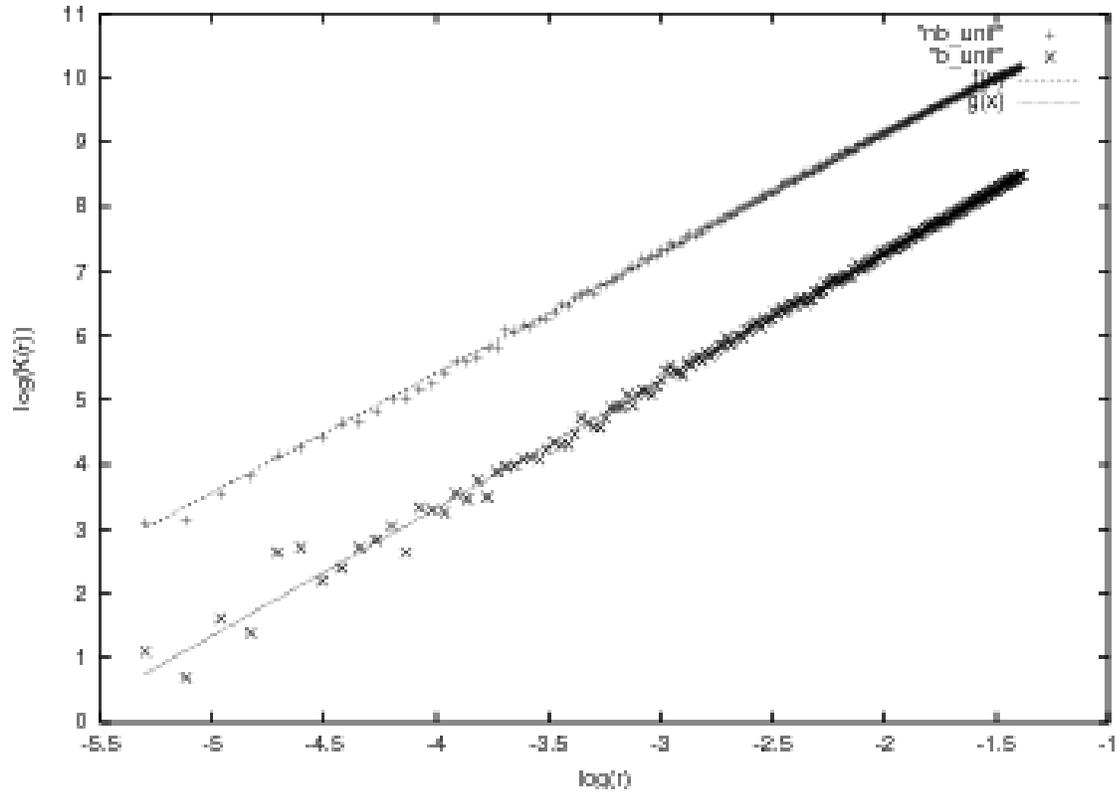}
\caption{Correlation histogram of an uniform gas in a unit cube, taking into account the edge shift (lower straight line, $D=3.00$, histogram accumulated less points), and without edge shift (upper curve, straight line adjustment gives $D=2.85$ -- obviously wrong). Notice that the correlation range is only 0.25. A little difference  in view, but great in numbers.} \end{figure}

\begin{figure}  
\includegraphics[scale=0.5, viewport=300 200  1100 900, clip=true]{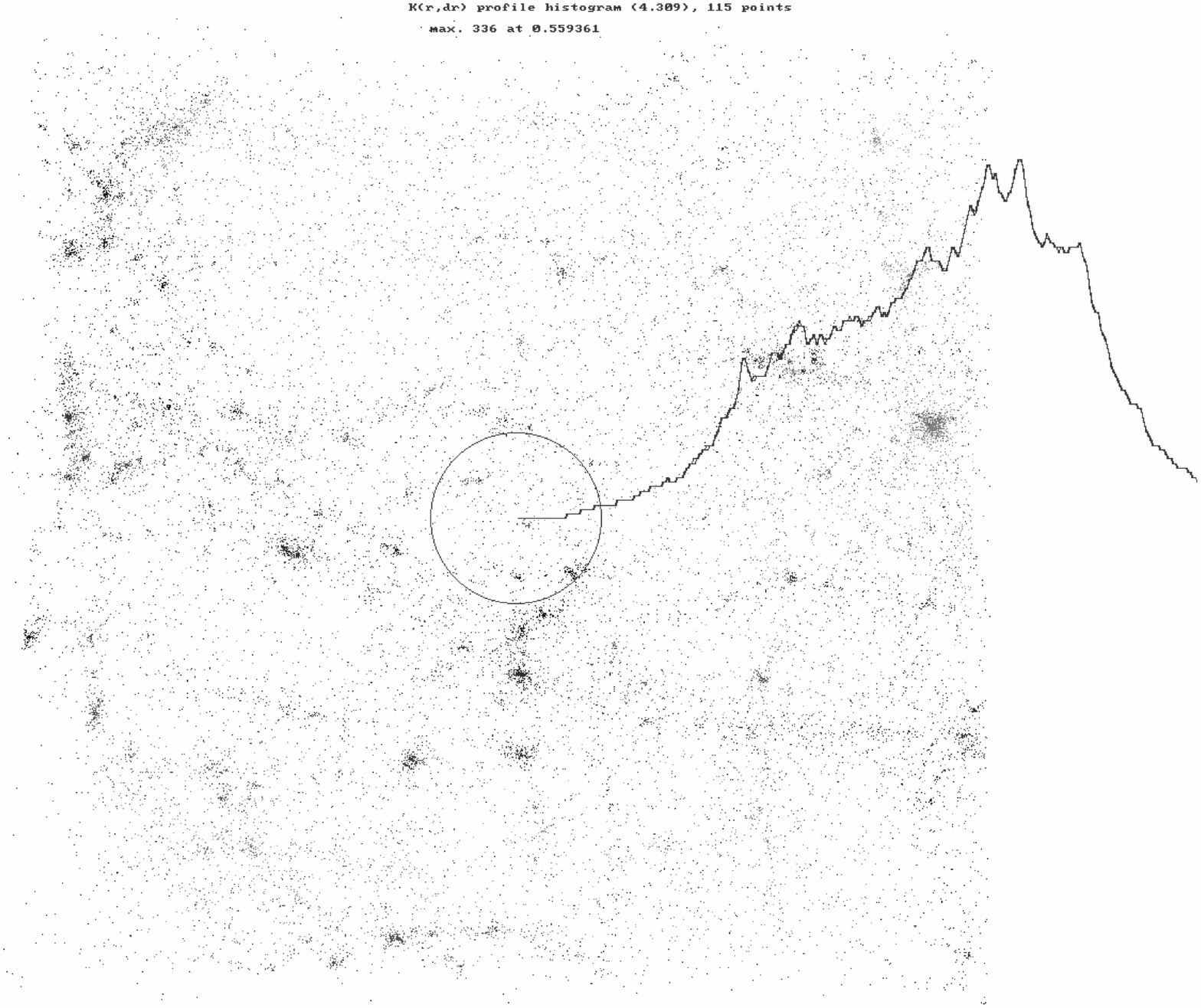}
\caption{Behavior of  the  $K$-estimator on a \textit{Virgo project} simulation sample -- the trend resembles a parabola, and the distribution histogram  is gas-alike, at least locally around  an initial point -- this distribution looks  almost self-similar, but with $D \approx 2.63$, not $D \approx 2$ as expected; we also see the edge influence on the right hand side. Using \textit{RoPo}, we can move fast through a given distribution from one point to another, and on each position instantly find a histogram.}		\end{figure}

\begin{figure} \includegraphics[scale=0.6,viewport=400 300  880  724,clip=true]{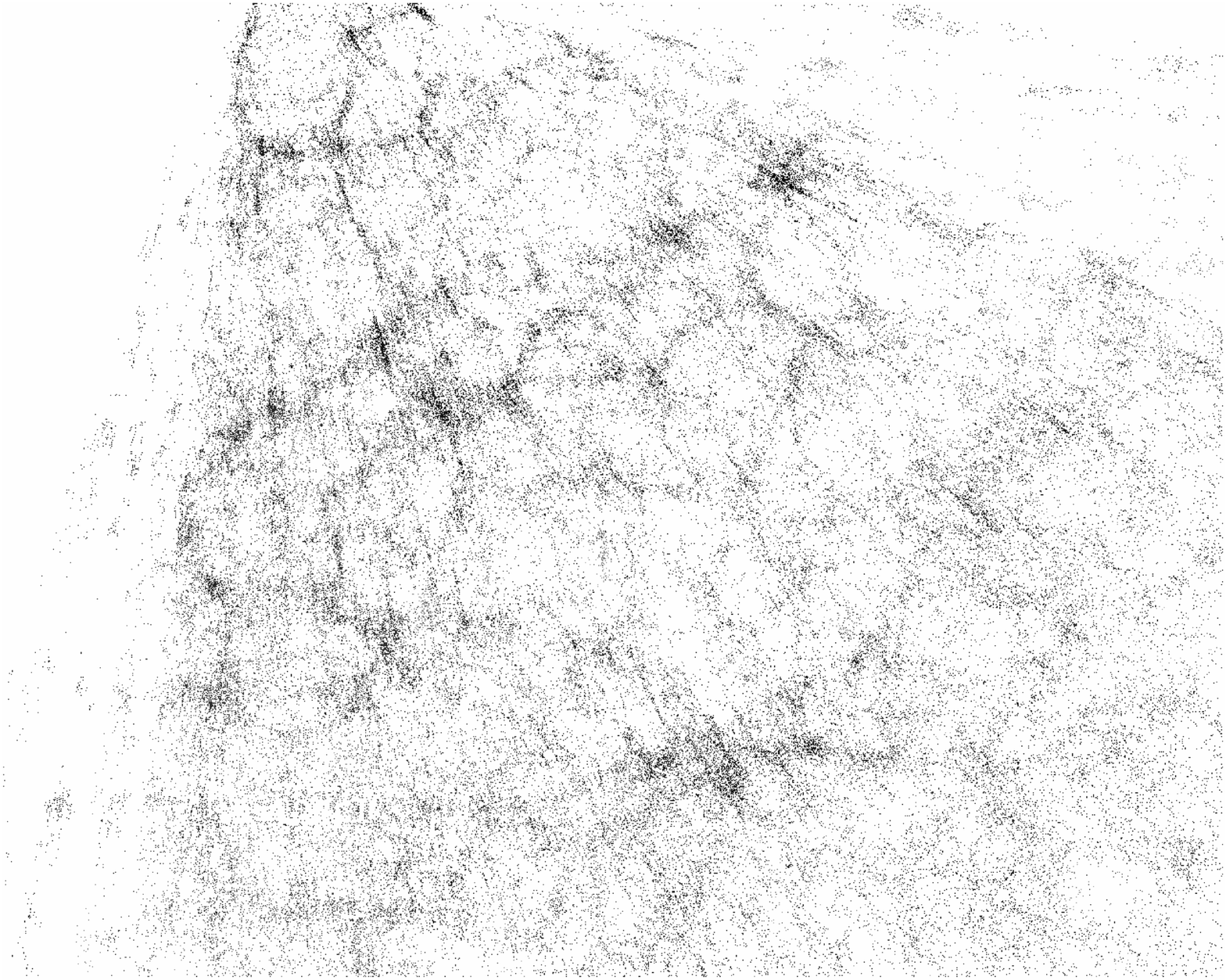} 
\caption{Clustering in 2dfGRS data; looks like points are running away from voids and this is a discrepancy with simulations data where points tend to collapse - cores can be clearly distinguished, see next figure.}   
\end{figure}

\begin{figure} \includegraphics[scale=0.6,viewport=400 300  880  724,clip=true]{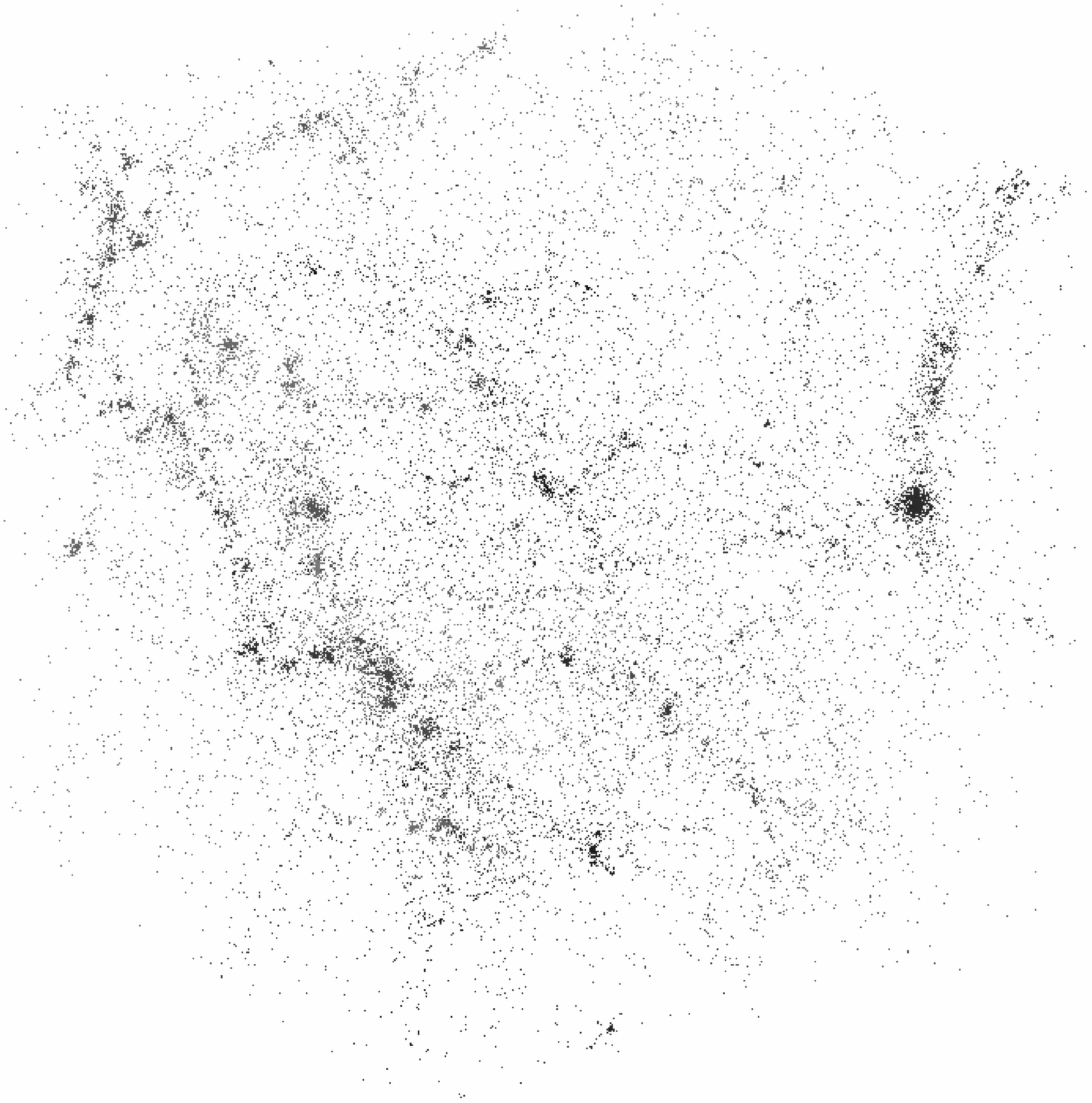} 
\caption{Simulations -- by correlation analysis more like a gas than 2dfGRS. If we replace clusters with points -- they are even more like a gas.}
\end{figure}

\begin{figure}   \includegraphics[scale=0.6,viewport=400 300  880  724,clip=true]{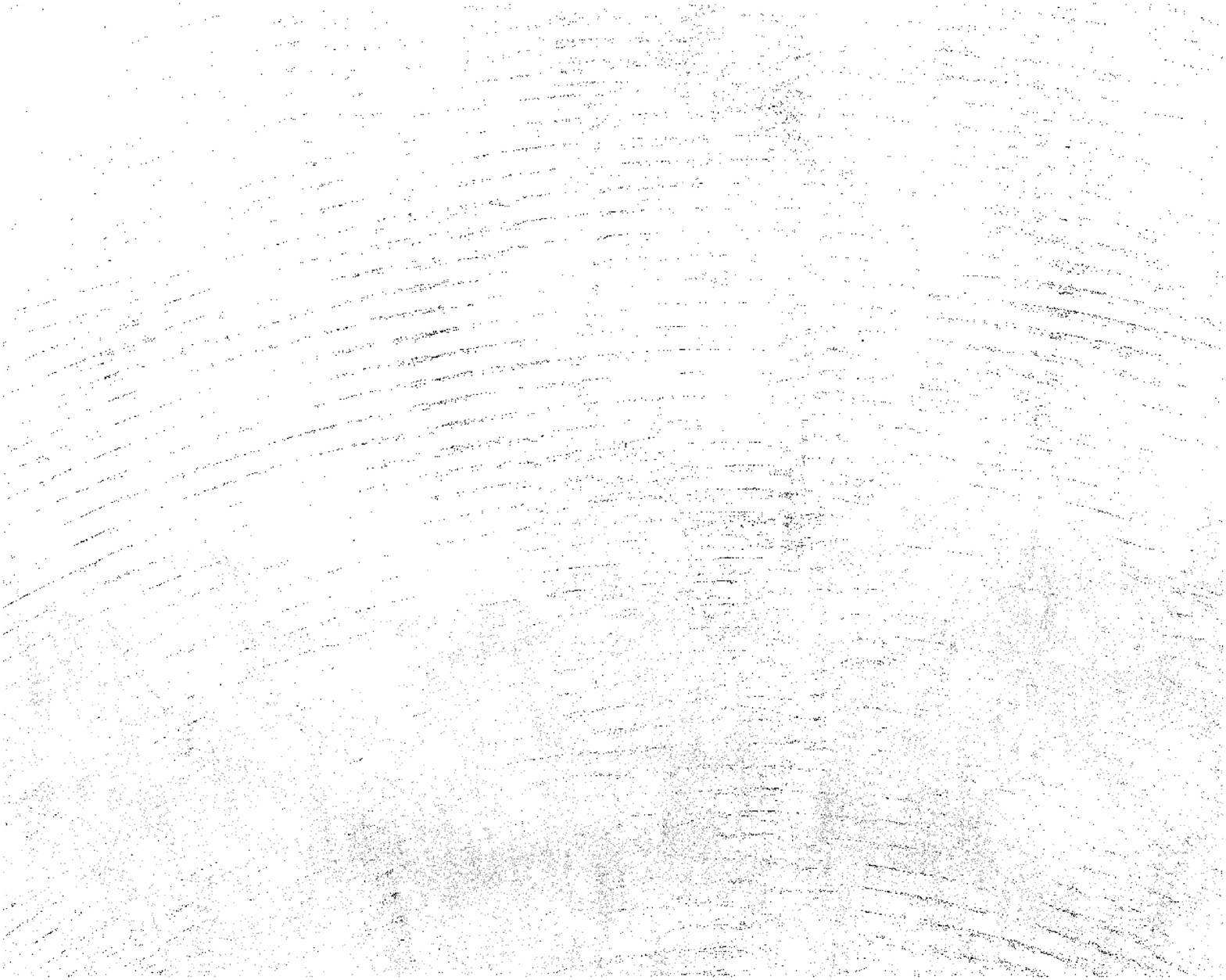} 
\caption{SDSS data flaw -- red shifts are rounded at a third decimal point and as we can take only small samples we observe slices - estimator clearly recognizes this artificial effect giving  $D \approx 1.8$  instead of expected $D \approx 2$.}          
\end{figure}

\begin{figure}
\includegraphics[scale=0.5]{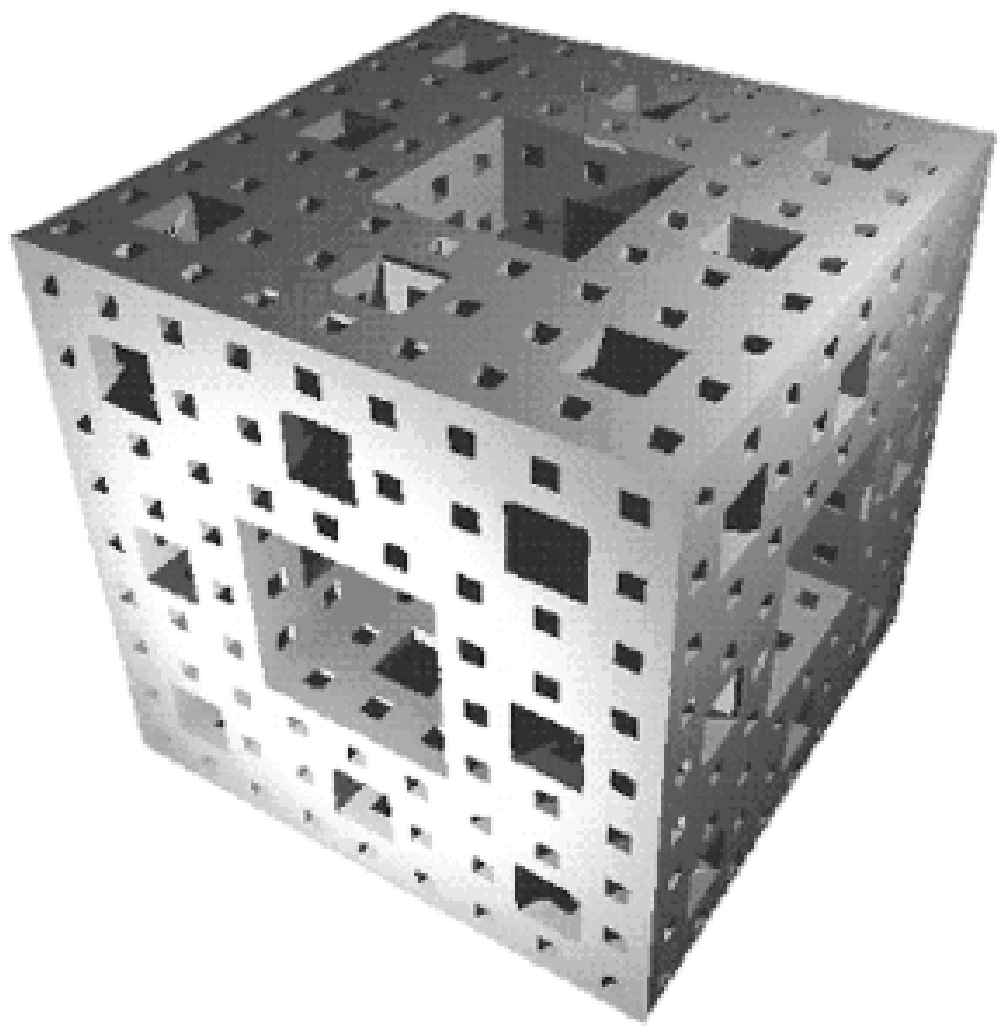}	 \caption{Menger sponge - picture taken from www.sewanee.edu/physics/ PHYSICS123/ physics123.html}
\end{figure} 

{\small  The main part of Korela procedure, data rescaled on to unit cube, Mersenne twister used for a greater speed:
\begin{verbatim}
while (n1<200000&&np<5000000){np=np+1;
/* first point */  i=(long)(n*genrand_real2()); /* RNG */
t0=t[i][0];if(t0<0.25||t0>0.75)goto jump;  
t1=t[i][1];if(t1<0.25||t1>0.75)goto jump; 
t2=t[i][2];if(t2<0.25||t2>0.75)goto jump;
/* second point */ j=(long)(n*genrand_real2());  
k0=t0-t[j][0];k1=t1-t[j][1];k2=t2-t[j][2];
k=(long)(1000.*sqrt(k0*k0+k1*k1+k2*k2)); // index
if((k>250)||(k<5))goto jump; u1[k]=u1[k]+1;n1=n1+1; jump:;}\end{verbatim} }

\end{document}